\documentclass[aps,preprint,prd,showpacs,nofootinbib]{revtex4-1}
\pdfoutput=1
\usepackage{graphicx}
\usepackage{longtable}
\usepackage{float}
\usepackage{dcolumn}
\usepackage{bm}
\usepackage{appendix}
\usepackage{multirow}
\usepackage{color}
\usepackage[utf8]{inputenc}
\usepackage{footmisc}
\usepackage{soul}
\usepackage{txfonts}
\usepackage{xcolor}
\usepackage{graphicx}
\usepackage{bm}
\usepackage{ulem}
\usepackage{multirow, array, float, tabularx}

\usepackage{amsmath}

\usepackage[colorlinks=true,linkcolor=blue,citecolor=blue]{hyperref}
\usepackage{feynmp-auto}
\usepackage{fancyhdr}
\usepackage{lineno}

\usepackage{wasysym}

\begin{document}
\title{Leptoquark-induced radiative masses for  active and sterile neutrinos  within the framework of the 3-3-1  model }

\author{A. Doff$^a$}
\author{Jo\~ao Paulo Pinheiro$^b$} 
\author{C. A. de S. Pires$^{c}$}
\affiliation{$^{a}$ Universidade Tecnologica Federal do Parana - UTFPR - DAFIS, R. Doutor Washington Subtil Chueire, 330 - Jardim Carvalho,  84017-220, Ponta Grossa, PR, Brazil} 
\affiliation{$^{b}$Departament de F\'isica Qu\`antica i Astrof\'isica and Institut de Ci\`encies del Cosmos, Universitat de Barcelona,
  Diagonal 647, E-08028 Barcelona, Spain}
  \affiliation{$^{c}$Departamento de F\'isica, Universidade Federal da Para\'iba, Caixa Postal 5008, 58051-970, Jo\~ao Pessoa, PB, Brazil}

\date{\today}
\vspace{1cm}
\begin{abstract}
In this work, we introduce the minimal set of leptoquarks into the 3-3-1 model with right-handed neutrinos, capable of generating radiative masses for active neutrinos. As a main consequence, the standard neutrinos acquire small Majorana masses at the one-loop level, while right-handed (sterile) neutrinos obtain small Majorana masses at the two-loop level, naturally making them light particles as well. Additionally, we discuss the viability of this scenario and several other interesting phenomenological consequences, including its impact on $B$-meson physics and rare Higgs decays, both of which are also induced by the leptoquarks.

\end{abstract}

\maketitle
\section{Introduction}
Although solar and atmospheric neutrino oscillation experiments revealed that neutrinos are massive particles  two decades ago\cite{Kajita:2016cak,McDonald:2019hkd}, the problem of the smallness of the neutrino masses remains an open question in theoretical particle physics\cite{ParticleDataGroup:2024cfk,Esteban:2024eli,Thorne:2024vsq, Altarelli:2002hx}. Such smallness is often explained through the dimension-5 Weinberg operator \cite{Weinberg:1979sa}. One common realization of this operator involves introducing a new scalar field that interacts with neutrinos and other particles, generating radiative neutrino masses via loop-level contributions \cite{Zee:1980ai,Babu:1988ki,CepedelloPerez:2021tgj,Cai:2017jrq}. Such mechanisms are more intricate in the context of UV-complete models, where the particle content tends to be large and diverse. When right-handed neutrinos are included, there are typically profound connections between active and sterile neutrino states. Understanding these connections is a key goal in studying UV-complete realizations of neutrino mass mechanisms \cite{Acero:2022wqg}.

Sterile neutrinos are hypothetical particles closely associated with active neutrinos. Despite their elusive nature, there are potential hints of their existence, providing significant motivation for their study. Several experimental anomalies can be better explained by the existence of sterile neutrinos that mix non-negligibly with active neutrinos, with masses in the eV scale. These anomalies include the reactor anomaly \cite{Mention:2011rk}, the gallium anomaly \cite{Hampel:1998xg,Kaether:2010ag}, the MiniBooNE anomaly \cite{Aguilar-Arevalo:2018gpe}, and the LSND anomaly \cite{Aguilar:2001ty}. All these short-baseline experiments exhibit tensions with the standard $3\nu$ scenario that are naturally explained by the presence of light sterile neutrinos \cite{Kopp:2013vaa}, which affect the oscillation probabilities of active neutrinos \cite{Gariazzo:2017fdh,Diaz:2019fwt}.

In addition to eV-scale sterile neutrinos, keV-scale sterile neutrinos are highly motivated in the context of dark matter studies \cite{Abazajian:2017tcc,Adhikari:2016bei}. They are a natural candidate for warm dark matter and could significantly influence the effective number of relativistic species during Big Bang Nucleosynthesis (BBN) \cite{Boehm:2012gr,Heeck:2012bq}. However, while light sterile neutrinos are theoretically appealing \cite{Dasgupta:2021ies,Abazajian:2017tcc,Drewes:2016upu,Boyarsky:2009ix}, generating small sterile neutrino masses while maintaining light the active neutrinos is a nontrivial challenge.

In the context of the $SU(3)_C \times SU(3)_L \times U(1)_N$, when right-handed neutrinos  appear in the fermion triplet together with active neutrinos and charged leptons, the variant is known as  331 with right-handed neutrinos (331RHN)\cite{Foot:1994ym,Montero:1992jk}. In its original version,  neutrinos are massless particles. The challenge here concerning neutrino mass generation is twofold: to explain active masses and understand its consequences on sterile neutrinos. There are few attempts in the literature investigating the generation of light sterile neutrinos in the 331RHN\cite{Dias:2005yh, Cogollo:2009yi,Cabrera:2023rcy}, here, for the first time, we address this issue by adding  leptoquarks to the 331RHN with the aim of inducing radiative masses for both active and sterile neutrinos\footnote{Leptoquarks were firstly introduced in the framework of 3-3-1 model in Ref. \cite{Doff:2023bgy}. In the particular framework of the 331RHN leptoquarks were firstly introduced in Ref. \cite{Doff:2024cap}}.

In this work, we show that using a minimal content of leptoquarks to generate active neutrino masses , composed by one triplet and one singlet scalars,   implies in generating light masses for the sterile states. As an original result, we show that active neutrinos acquire masses through  one-loop and sterile neutrinos through two-loops, making sterile neutrinos, unavoidably, a light particle. To complete our study, we verify the implications of active neutrino mass and mixing, for different orderings, with some rare $B$- meson decays and  the rare Higgs decay  $h \to \mu^+ \tau^-$.

This paper is organized in the following way: In Sec. (II) we present the essence of the model including leptoquarks. In Sec. (III) we investigate the features of the loops that generate neutrino masses. In Sec. (IV) and (V)  we investigate if light neutrinos are consistent with constraints from B-meson and rare Higgs decay physics. In Sec. (VI) we engage in discussions and  conclude  with a brief summary.

\section{The essence of our Framework}
The leptonic sector of the 331RHN is composed by  the standard leptons and right-handed neutrinos. This content  is arranged into triplet and singlet according to the 3-3-1 symmetry transforming as
 \cite{Foot:1994ym,Montero:1992jk},
\begin{equation}
L_{a_L}= \begin{pmatrix}
\nu_{a}     \\
e_{a}       \\
\nu^{c}_{a} \\
\end{pmatrix}_{L} \sim (1,3,-1/3), \quad e_{a_R}\sim (1,1,-1),
\label{LT}
\end{equation}
with $a=1\,,\,2\,\,3$ representing the three standard  generations of leptons.

The hadronic sector is significantly different from the standard model\cite{Foot:1994ym,Montero:1992jk}. To achieve anomaly cancellation one family of quarks must transform differently from the other two. In this work we have chosen the first two families of quarks  to transform as anti-triplet while the third one  transforming as triplet under  the 3-3-1 symmetry\footnote{For the other variants, see Refs.\cite{Oliveira:2022vjo}},
\begin{eqnarray}
&&Q_{i_L} = \left (
\begin{array}{c}
d_{i} \\
-u_{i} \\
d^{\prime}_{i}
\end{array}
\right )_L\sim(3\,,\,\bar{3}\,,\,0)\,,u_{iR}\,\sim(3,1,2/3),\,\,\,\nonumber \\
&&\,\,d_{iR}\,\sim(3,1,-1/3)\,,\,\,\,\, d^{\prime}_{iR}\,\sim(3,1,-1/3),\nonumber \\
&&Q_{3L} = \left (
\begin{array}{c}
u_{3} \\
d_{3} \\
u^{\prime}_{3}
\end{array}
\right )_L\sim(3\,,\,3\,,\,1/3),u_{3R}\,\sim(3,1,2/3),\nonumber \\
&&\,\,d_{3R}\,\sim(3,1,-1/3)\,,\,u^{\prime}_{3R}\,\sim(3,1,2/3),
\label{quarks} 
\end{eqnarray}
where  the index $i=1,2$ is restricted to only two generations. The negative signal in the anti-triplet $Q_{i_L}$ is  to standardize the signals of the charged current interactions with the gauge bosons.  The primed quarks are new heavy quarks with the usual $(+\frac{2}{3}, -\frac{1}{3})$ electric charges. 

The original scalar sector of the  model is composed by three triplets of scalars\cite{Foot:1994ym,Montero:1992jk}
\begin{eqnarray}
\eta = \left (
\begin{array}{c}
\eta^0 \\
\eta^- \\
\eta^{\prime 0}
\end{array}
\right ),\,\rho = \left (
\begin{array}{c}
\rho^+ \\
\rho^0 \\
\rho^{\prime +}
\end{array}
\right ),\,
\chi = \left (
\begin{array}{c}
\chi^0 \\
\chi^{-} \\
\chi^{\prime 0}
\end{array}
\right ),
\label{scalarcont} 
\end{eqnarray}
with $\eta$ and $\chi$ transforming as $(1\,,\,3\,,\,-1/3)$
and $\rho$ as $(1\,,\,3\,,\,2/3)$. 

The minimal content of  leptoquarks required to achieve our goal include one triplet and one singlet of scalar leptoquarks transforming as, 
\begin{eqnarray}
T = \left (
\begin{array}{c}
T^{+1/3} \\
T^{-2/3} \\
T^{\prime +1/3}
\end{array}
\right )\sim (3\,,\,3\,,\,0),\,\,\,\,\,\,S^{+1/3}\sim(3\,,\,1\,,\,1/3)
\end{eqnarray}

To work with the simplest scenario, we impose the lagrangian of the model to be symmetric by a $Z_2$ symmetry with  $(T\,,\,\eta\,,\,\rho\,,\,e_{a_R}\,,\,u_{a_R}\,,\,d_{a_R}) \to -(T\,,\,\eta\,,\,\rho\,,\,e_{a_R}\,,\,u_{a_R}\,,\,d_{a_R}) $ while all the other fields transform trivially by the $Z_2$ symmetry.

With all this in hand,  the  Yukawa interactions among fermions and scalars in the model are composed by the following terms,
\begin{eqnarray}
-{\cal L}_Y &=&  \tilde g_{ij} \bar Q_{i_L} \chi^* d^{\prime}_{j_R} + \tilde h_{33} \bar Q_{3_L} \eta u^{\prime}_{3_R} + g_{ia} \bar Q_{i_L} \eta^* d_{a_R} + h_{3a} \bar Q_{3_L} \eta u_{a_R} + g_{3a} \bar Q_{3_L} \rho d_{a_R} +\nonumber \\
&+& h_{ia} \bar Q_{i_L} \rho^* u_{a_R} + Y_{ab}\bar L_{a_L} T d_{b_R} +\tilde{Y}_{ia}\bar Q^C_{i_L} L_{a_L}S  + y_{ab}\bar u^C_{a_R} e_{b_R} S +f_{aa}\bar L_{a_L}\rho e_{a_R}   +\mbox{H.c.}\,,
\label{yukawa}
\end{eqnarray}
where $a=1,2,3$ and $i=1,2$.

The potential of the model is composed by the terms
\begin{eqnarray}
V(\eta\,,\,\rho\,,\,\chi\,,\,T\,,\,S)&=&V(\eta\,,\,\rho\,,\,\chi)+\mu^2_{T}T^{\dagger}T  +\mu^2_S S^{\dagger}S+\tilde \lambda_1|T^{\dagger} T|^2+ \tilde \lambda_2 |S^{\dagger}S|^2 +\nonumber \\
&&(\tilde \lambda_3 \eta^{\dagger}\eta+\tilde \lambda_4\rho^{\dagger}\rho+ \tilde \lambda_5 \chi^{\dagger}\chi)T^{\dagger} T +(\tilde \lambda_6 \eta^{\dagger}\eta+\tilde \lambda_7\rho^{\dagger}\rho+ \tilde \lambda_8 \chi^{\dagger}\chi)S^{\dagger} S +\nonumber \\
&&\tilde \lambda_9 |\eta^{\dagger} T|^2 +\tilde \lambda_{10} |\rho^{\dagger} T|^2+\tilde \lambda_{11} |\chi^{\dagger} T|^2+ \tilde \lambda_{12} T^{\dagger} TS^{\dagger} S +(M T^{\dagger}\eta S + H.c.),
\label{VSLQ}
\end{eqnarray}
with $V(\eta\,,\,\rho\,,\,\chi)$ being  the original scalar potential given in Refs. \cite{Pal:1994ba, Long:1997vbr,Ponce:2002sg,Pinheiro:2022bcs} while the parameter $M$ is the energy scale associated to the explicit violation of lepton number. After spontaneous breaking of the  3-3-1  symmetry\footnote{ $ \langle \eta^0 \rangle =v_\eta\,\,,\,\, \langle \rho^0 \rangle =v_\rho\,\,,\,\,\langle \chi^{\prime 0} \rangle =v_{\chi^{\prime}}$ with $v_{\chi^{\prime}} \gg v_\eta\,,\,v_\rho $ and $v^2_\eta + v^2_\rho=v^2_{ew}$ },  the resulting scalar spectrum  involves  five scalars $(h_1\,,\,h_2\,,\,h^{\pm}_1\,,\,A)$ with mass at the electroweak scale  and other five scalars $(H\,,\,h_2^{\pm}\,,\,\chi^0 \,,\, \eta^{\prime 0})$  with mass belonging to 3-3-1 scale\cite{Pinheiro:2022bcs,Long:1997vbr}.

When  $\eta^0$ develops VEV the last term in the potential promotes a mixing among the leptoquarks $T^{+1/3}$ and $S^{+1/3}$. Considering  the basis $(S^{+1/3}\,,\,T^{+1/3})^T$ we get

\begin{equation} 
M^2_{1/3}\approx \left(\begin{array}{cc}
 \mu_S^2 + \frac{1}{2}(\tilde \lambda_6 v^2_\eta + \tilde \lambda_7 v^2_\rho + \tilde 
 \lambda_8 v^2_{\chi^{\prime}}) & \frac{M v_\eta}{\sqrt{2}}\\ 
 \frac{M v_\eta}{\sqrt{2}} &  \mu_T^2 +\frac{1}{2}\left( (\tilde \lambda_3 +\tilde \lambda_9)v^2_\eta + (\tilde \lambda_4 + \tilde \lambda_{10}) v^2_\rho + (\tilde 
 \lambda_5 + \tilde \lambda_{11}) v^2_{\chi^{\prime}})\right)
      \end{array}\right).
      \label{matrixfinal2}
\end{equation}
After  diagonalize this mass matrix we obtain the eigenvalue $M_1$ associated to $ \tilde S^{+1/3}$ and the eigenvalue $M_2$ associated to $\tilde T^{+1/3}$. The relation among the basis is given by 
\begin{eqnarray}
\left (
\begin{array}{c}
\tilde S^{+1/3} \\
\tilde T^{+1/3} 
\end{array}
\right )= 
\left(\begin{array}{cc}
 C_\theta & -S_\theta\\ 
 S_\theta &  C_\theta
      \end{array}
    \right)
   \left (
   \begin{array}{c}
 S^{+1/3} \\
 T^{+1/3} 
\end{array}
\right),
\label{RLQ}
\end{eqnarray}
 with 
\begin{equation}
    \sin2 \theta=\sqrt{2} \frac{Mv_\eta}{M^2_1-M^2_2}.
    \label{2theta}
\end{equation}

The gauge sector of the model involves the standard gauge bosons plus one neutral gauge boson $Z^{\prime}$, two new charged gauge bosons $W^{\prime \pm}$ and two non-hermitian neutral gauge bosons $U^0$and $U^{0 \dagger}$\cite{Long:1995ctv}. All the new gauge bosons  develop mass at $v_{\chi^{\prime}}$ scale.


\section{neutrinos masses}
In the original version of the 331RHN, neutrinos are massless \cite{Foot:1994ym,Montero:1992jk,Long:1995ctv}. A very versatile way of generating masses for the neutrinos is by introducing a sextet of scalars to the original scalar content of the model. The sextet can enable the canonical type-I and type-II seesaw mechanisms \cite{Ky:2005yq,Cogollo:2008zc,Dong:2008sw,deSousaPires:2018fnl}. Another possibility arises with the addition of new neutral chiral fermions, which can facilitate the inverse seesaw mechanism \cite{Dias:2012xp}. Radiative mass generation has also been explored in Refs.~\cite{Boucenna:2014ela,CarcamoHernandez:2021tlv,Das:2020pai,Binh:2024lez}. 

In most cases, right-handed neutrinos acquire masses significantly heavier than left-handed neutrinos. This result is expected since right-handed neutrinos are singlets under the Standard Model gauge symmetry. However, it is possible to arrange the model such that right-handed neutrinos can acquire light masses \cite{Dias:2005yh,Cogollo:2009yi,Cabrera:2023rcy}.

In this work, we propose and develop a novel mechanism for generating tiny neutrino masses within the 331RHN model, leveraging radiative corrections induced by leptoquarks. The relevant terms in the Yukawa Lagrangian, which contribute to neutrino mass generation, are proportional to the couplings $Y$ and $\tilde{Y}$. By expanding these terms, we obtain

\begin{eqnarray}
 \bar Q^C_{i_L} L_{a_L}S \supset  \bar d_{i_L}  (\nu_{a_L})^C S^* +\bar d^{\prime }_{i_L} \nu_{a_R} S^* &=& C_\theta \bar{\hat{d}}_{b_L}(V^d_L)^*_{bi} ( \nu_{a_L})^C \tilde S^{-1/3}+S_\theta \bar{\hat{d}}_{b_L}(V^d_L)^*_{bi} ( \nu_{a_L})^C \tilde{T}^{-1/3}+\nonumber \\
 &&+ C_\theta \bar{\hat{d^{\prime}}}_{i_L}  \nu_{a_R} \tilde S^{-1/3}+S_\theta \bar{\hat{d^{\prime}}}_{i_L}  \nu_{a_R} \tilde{T}^{-1/3}
 \label{LQ1}
\end{eqnarray}
and 
 \begin{eqnarray}
  \bar L_{a_L} T d_{b_R} \supset \bar d_{b_R} \nu_{a_L}T^{-1/3}+\bar d_{b_R} (\nu_{a_R})^C T^{\prime -1/3}=C_\theta \bar{\hat{d}}_{b_R}\nu_{a_L}\tilde{T}^{-1/3} -
  S_\theta \bar{\hat{d}}_{b_R}\nu_{a_L}\tilde{S}^{-1/3} + \bar{\hat{d}}_{b_R}(\nu_{a_R})^C\tilde{T^{\prime}}^{-1/3},
  \label{LQ2}
\end{eqnarray}
\begin{figure}[t]
\centering
\hspace*{-1cm}\includegraphics[width=1\columnwidth]{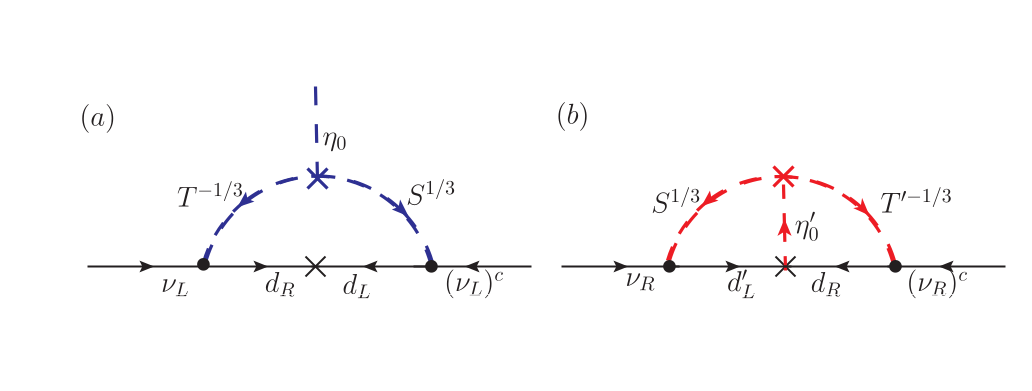}
\vspace*{-1.5cm}
\caption{ The  left-handed neutrinos mass at one-loop (a) and right-handed neutrinos mass at two-loop (b).}
\label{a331rhn}
\end{figure}
where  $\theta$ is the angle that  transform  $T^{1/3}$ and $S^{1/3}$  in the physical eigenstates $\tilde {T}^{1/3}$ and $\tilde {S}^{1/3}$.  As we assumed that  $\eta^{\prime 0}$ is inert,  consequently  $T^{\prime 1/3}=\tilde {T}^{\prime 1/3}$,
furthermore, let us consider $M_{T^{\prime 1/3}} = M_2$.

The mixing matrix $V^d_L$ transforms the flavor eigenstate $d_L$ in the physical ones $\hat d_L$. Moreover they must respect the constraints $V^{\dagger d}_L V^u_L=V_{CKM}$. We also  assume that  $d_R$, $u_R$ and $d^{\prime}_{L,R}$ come in a diagonal basis. All this are common assumptions.

In this point we stress that  the term $M T^{\dagger} \eta S^{1/3}$ in the potential in Eq. (\ref{VSLQ}) is essential in obtaining neutrino masses since it violates lepton number by two units.

Before delving  into the details,  we note that the interactions described above  lead to a very original result: left-handed neutrinos will  acquire masses at the one-loop level, while right-handed neutrinos will acquire masses at the two-loops level.  This  predicts that  sterile neutrinos are inexorably  light particles. The diagrams in the  flavor basis illustrating  the  one-loop and two-loops processes  generating neutrino masses  are shown in FIG.\ref{a331rhn}. We will develop each of these terms in detail  below.

\subsection{One-loop  active neutrinos masses}
\par The Yukawa interactions in Eqs. (\ref{LQ1}) and (\ref{LQ2}) can generate masses for the active neutrinos at the one-loop level, mediated by the leptoquarks $\hat{T}$ and $\hat{S}$, as shown in FIG. \ref{a331rhn}(a). The resulting active neutrino mass matrix in the flavor basis is expressed as \cite{Kova:Kova07}\cite{Chen_2022}\cite{Babu:2020hun}
\begin{equation}
    (m_\nu)^L_{ab} = \frac{3}{16\pi^2}\sum_{i=1,2}\sum_{j=d,s,b}m_j B_{0}(0,m^2_j,M^2_{i})\left(U_{i1}(\theta)U_{i2}(\theta)\left[ (V^{d*}_L \tilde{Y})^{aj}Y^{jb} + (V^{d*}_L  \tilde{Y})^{bj}Y^{ja}\right] \right),
    \label{Mnu}
\end{equation}
where $M^2_{i}$ ($i=1,2$) are the mass eigenvalues of the leptoquarks, 
$U_{ij}(\theta)$ is the rotation matrix that diagonalizes the leptoquark mass basis, as defined in Eq.(\ref{matrixfinal2}), and $B_{0}(0,m^2_j,M^2_i)$ is the Passarino-Veltman function, given by 
\begin{equation}
    B_{0}(0,m^2_j,M^2_i) = \frac{m^2_j\ln(m^2_j)-M^2_i\ln(M^2_i)}{m^2_j - M^2_i}.
    \label{B00PV}
\end{equation}

The matrix $(m_\nu)^L_{ab}$ must be consistent with the current constraints on the neutrino parameters \cite{Esteban:2024eli}.  Neglecting the CP phase of the neutrinos, the neutrino sector has five free parameters: three mixing angles, $\theta_{12}$, $\theta_{13}$, and $\theta_{23}$, as well as two squared mass differences, $\Delta m_{21}^2$ and $\Delta m_{31}^2$, which are required by solar and atmospheric neutrino oscillations. Imposing  that $(m_\nu)^L_{ab}$, after diagonalized, recovers the current values of   $\theta_{12}$, $\theta_{13}$, $\theta_{23}$, $\Delta m_{21}^2$ and $\Delta m_{31}^2$ translates into a set of conditions on the parameters that define $(m_\nu)^L_{ab}$. Note that $(m_\nu)^L_{ab}$ given above involves much  more free parameters than the necessary to recover uniquely the five free parameters responsible by the neutrino sector. Assuming that $m_{d,s,b}$ are much smaller than $M_{1,2}$ and that the up quarks come in a diagonal basis, which means $V^d_L=V_{CKM}$, the  Eq. (\ref{Mnu}) is opened in the following  terms:
\begin{eqnarray}
    (m_{\nu})^L_{ab} = \frac{3}{32\pi^2}\sin 2\theta \ln{\frac{M^2_1}{M^2_2}}\sum_p\Bigg(&& m_d[(V_{\mathrm{CKM}})_{ap} \tilde{Y}_{p1}Y_{1b} + ( V_{\mathrm{CKM}})_{bp}\tilde{Y}_{p1}Y_{1a}] +\nonumber \\
    && m_s[(V_{\mathrm{CKM}})_{ap} \tilde{Y}_{p2}Y_{2b} + ( V_{\mathrm{CKM}})_{bp}\tilde{Y}_{p2}Y_{2p}]+\nonumber \\
    && m_b[(V_{\mathrm{CKM}})_{ap} \tilde{Y}_{p3}Y_{3b} + ( V_{\mathrm{CKM}})_{bp}\tilde{Y}_{p3}Y_{3a}]\Bigg),
    \label{Mnu2}
\end{eqnarray}

In this work, we will study separately Normal Ordering (NO) and Inverted Ordering (IO) in order to investigate the impact of the neutrino mass ordering in the hadronic sector. For both orderings, the mixing angles $\theta_{12}$, $\theta_{13}$ and the mass squared differences $\Delta m_{\mathrm{solar}}^2$, $\Delta m_\mathrm{atm}^2$ are fixed by the bestfit values of NuFIT\cite{Esteban:2024eli}, while we are not adopting a preference for an octant of $\theta_{23}$ ($\sin^2\theta_{23}=0.5$) and fixing to zero the CP phase ($\delta_\mathrm{CP}$). 

For the oscillation data it is not possible to obtain the absolute mass of neutrinos, only the mass differences. Then, we are letting  $m_1(m_3)= 0$, for NO(IO). For this choice, for NO, $m_2= \sqrt{\Delta m_{\mathrm{solar}}^2}$ and $m_3 = \sqrt{\Delta m_{\mathrm{atm}^2}}$, while for the IO, $m_2= \sqrt{\Delta m_{\mathrm{solar}}^2}+\sqrt{\Delta m_{\mathrm{atm}}^2}$ and $m_1 = \sqrt{\Delta m_{\mathrm{atm}}^2}$, which implies that the heaviest active neutrino has mass $\approx 0.5$ eV. To estimate the order of the Yukawa matrices $\tilde{Y}$ and $Y$, we impose that Eq. (\ref{Mnu2}) recover the full neutrino mass matrix in flavor basis.  Then, we are solving a system with 6 equations with 12 variables (6 real variables for the components of each Yukawa matrix $\tilde{Y}$ and $Y$). Assuming the above described values for the neutrino parameters and the quark masses and the CKM matrix from PDG\cite{ParticleDataGroup:2024cfk} we solve the system 
\begin{equation} 
{(m_\nu)^L_{ab}}^{EXP} ={(m_\nu)^L_{ab}}^{PRED}      \label{generalMnu}
\end{equation}
where ${(m_\nu)^L_{ab}}^{EXP}=U^T_{PMNS} (m_\nu)^D U_{PMNS}$  with $m^D_\nu=diag(m_1\,\, m_2\,,\,m_3)$. In this case  ${(m_\nu)^L_{ab}}^{EXP}$ re\-pre\-sen\-ts the experimental value for the active neutrino mass matrix in flavor basis for each mass ordering  and ${(m_\nu)^L_{ab}}^{PRED}$ is the model prediction for the active neutrino mass at one loop. By solving the above sistem, it is possible to find the entries of the matrices $Y$ and $\tilde Y$. For the NO case we have:

\begin{equation}
   Y_\mathrm{NO}= \left(
\begin{array}{ccc}
 0.6728 & 0.1295 & -0.0031 \\
 0.1295 & 0.0236 & 0.0009 \\
 -0.0031 & 0.0009 & -0.0019 \\
\end{array}
\right),\,\,
 \tilde Y_\mathrm{NO} =  \left(
\begin{array}{ccc}
 0.8822 & 0.0220 & 0.2103 \\
 0.0220 & -0.0237 & 0.0926 \\
 0.2103 & 0.0926 & 0.0032 \\
\end{array}
\right),
\label{YNO}
\end{equation}
while for  IO case  we obtain:
\begin{equation}
   Y_\mathrm{IO}= \left(
\begin{array}{ccc}
 -0.6058 & -0.1263 & 0.0005 \\
 -0.1263 & -0.0370 & 0.0052 \\
 0.0005 & 0.0052 & 0.0017 \\
\end{array}
\right),\,\,
 \tilde Y_\mathrm{IO} = \left(
\begin{array}{ccc}
 -0.4446 & 0.0756 & 0.1300 \\
 0.0756 & 0.2273 & 0.0695 \\
 0.1300 & 0.0695 & -0.0021 \\
\end{array}
\right).
\label{YIO}
\end{equation}
In both cases the Yukawa couplings vary in the range $10^{-1}-10^{-4}$ which are plausible values for these couplings. In short, we have active neutrino with mass at eV scale for very plausible values for the parameters involved in the neutrino mass expression as leptoquarks at TeV scale and Yukawa couplings within a range of values compatible with the values of the standard Yukawa couplings of the charged fermions in the SM. In other words, our model provides eV neutrinos in a conservative way.

\subsection{Two-loop right-handed neutrinos masses}

The Yukawa interactions in Eqs. (\ref{yukawa}), (\ref{LQ1}) and (\ref{LQ2}) yield inevitably to two-loops  Majorana masses to the  right-handed neutrinos  as depicted in FIG.\ref{a331rhn}(b).  The loop is mediated by the standard singlet down quarks, $d_R$, the new quarks, $d_i^{\prime}$, the leptoquarks $S^{1/3}$ and $T^{\prime 1/3}$ and the scalar $\eta^{\prime 0}$. As $d_R$ we assume that  $d_i^{\prime}$ comes, too, in a diagonal basis. Thus, differently from the one loop case above of the standard neutrinos, the masses of the right-handed neutrinos will not involve the quarks mixing.

 The expression to the sterile neutrino masses that arises from the loops in FIG. \ref{a331rhn}(b) is found in Refs. \cite{Babu:2020hun,Babu:Babu88,McDonald:2003MK,Babu:2019mfe,Babu:2010vp}. The respective resulting sterile neutrino mass matrix in the flavor basis can be written as 

\begin{eqnarray}
&& (m_\nu)^R_{ab} = \frac{3 M}{(2\pi)^8}\sum_{i=1,2}\sum_{j=d,s,b}  m_j m_{d^\prime_i} U_{11}(\theta) h(m_{d^\prime_i}, M_{\eta^{\prime}_0},m_{j}, M_S, M_{T^\prime})\times\nonumber \\
&&\hspace*{4.6cm}\left[ \tilde{Y}^{ai}g^{ip}Y^{pb} + \tilde{Y}^{bi}g^{ip}Y^{pa}  \right]  + \nonumber \\
&&\hspace*{1.7cm} \frac{3 M}{(2\pi)^8}\sum_{i=1,2}\sum_{j=d,s,b} m_j m_{d^\prime_i} U_{12}(\theta) h(m_{d^\prime_i}, M_{\eta^{\prime}_0},m_{j}, M_S, M_{T^\prime})\times\nonumber\\
&&\hspace*{4.6cm}\left[ \tilde{Y}^{ai}g^{ip}Y^{pb} + \tilde{Y}^{bi}g^{ip}Y^{pa}  \right].
  \label{MnuR}
\end{eqnarray}
\noindent Before we proceed in determining $(m_\nu)^R$, let us make some observations: First, differently from the one-loop of the FIG. \ref{a331rhn}(a), in the two-loops in the FIG.\ref{a331rhn}(b) only right-handed standard quarks run into the loop which means we do not have the suppression factor due to quarks mixing $V^{u,d}_L$. In this expression the new quarks $d^{\prime}_{i_L}$ are naturally assumed in a diagonal basis. Note that as we commented below Eq.(\ref{yukawa}),  the index $i=1,2$ is restricted to only two quarks generations. The factor $M$ was introduced in Eq.(\ref{VSLQ}) and generates the mixing between leptoquarks $S^{1/3}$ and $T^{1/3}$ which trigger the masses of the neutrinos.
The leptoquark $T^{\prime 1/3}$ carries  331 degrees of freedom and does not mix with leptoquarks $S^{1/3}, T^{1/3}$,  as a consequence in Eq.(\ref{MnuR}), we have
that sterile neutrino masses are  proportional to $\cos\theta$(or $\sin\theta$) and not to  $\sin 2\theta$ as occurs at 1-loop level. In order to simplify even more our analysis we will assume that   $m_{d^\prime_i} \equiv m_{d^\prime}$, and  that all leptoquarks in the FIG.\ref{a331rhn}(b)
lies at the TeV scale, more precisely that these masses are of the same order as $M_2$. In this case  the integrals described in Eq.(\ref{MnuR}), 
can be written  as $h(m_{d^\prime}, M_{\eta^{\prime}_0}, m_{j},M_2)$  and we obtain 
\begin{eqnarray}
 (m_\nu)^R_{ab} = \frac{3 M  m_{d^\prime}}{(2\pi)^8}\sum_{i=1,2}\sum_{j=d,s,b} m_j h( m_{d^\prime}, M_{\eta^{\prime}_0},m_{j}, M_2)\left[ \tilde{Y}^{ai}g^{ip}Y^{pb} + \tilde{Y}^{bi}g^{ip}Y^{pa}  \right] .
  \label{MnuR2}
\end{eqnarray}
\par Since that all quarks in the loop  are in a diagonal basis, we have 
\begin{eqnarray}
&& \hspace*{-0.7cm}  (m_\nu)^R_{ab} = \frac{3 M m_{d^\prime}m_b}{(2\pi)^8}\sum_{i=1,2} \left(g_{i1}\frac{m_d}{m_b} h( m_{d^\prime}, M_{\eta^{\prime}_0},m_{d}, M_2)\left[ \tilde{Y}^{ai}Y^{1b} + \tilde{Y}^{bi}Y^{1a}  \right] + \right.\nonumber\\ 
&&\hspace*{4cm} \left. g_{i2}\frac{m_s}{m_b} h( m_{d^\prime}, M_{\eta^{\prime}_0},m_{s}, M_2)\left[ \tilde{Y}^{ai}Y^{2b} + \tilde{Y}^{bi}Y^{2a}  \right] + \right.\nonumber\\
&&\hspace*{4.6cm} \left. g_{i3} h(m_{d^\prime}, M_{\eta^{\prime}_0},m_{b}, M_2)\left[ \tilde{Y}^{ai}Y^{3b} + \tilde{Y}^{bi}Y^{3a} \right]\right)
  \label{MnuR3}
\end{eqnarray}
\par Therefore, assuming that $m_b \gg m_d, m_s$, the maximum contribution to $(m_\nu)^R_{ab}$ can be identified as
\begin{eqnarray}
 (m_\nu)^R_{ab} = \frac{3 M m_{d^\prime}m_b}{(2\pi)^8} h(m_{d^\prime}, M_{\eta^{\prime}_0},m_{b}, M_2)\sum_{i=1,2}g_{i3}\left[ \tilde{Y}^{ai}Y^{3b} + \tilde{Y}^{bi}Y^{3a} \right].
  \label{MnuR4}
\end{eqnarray}
\noindent such that $h( m_{d^\prime}, M_{\eta^{\prime}_0},m_{b}, M_2)$  is given in Appendix \ref{APA}. 

By fixing $M_2=2$ TeV, we obtain the value of $M=147$ GeV for $\sin2 \theta=10^{-2}$. For the other typical 3-3-1 particles, we fix $m_{d^\prime}= 1$ TeV and $M_{\eta^{\prime}_0}=3$ TeV. Moreover for $v_\eta=v_\rho$ we obtain
$g_{13} = 0.000119$  and  $g_{23} = 0.0014$. For these benchmark points, it is possible to estimate the value of the sterile neutrinos masses for NO and IO discussed above for the standard neutrinos. This is a peculiar feature of the 331RHN since left-handed and right-handed neutrinos compose the same lepton triplet then they share the same Yukawa interactions with the leptoquarks. In this way the Yukawa couplings that compose the mass matrix of the left-handed neutrinos will appear in the mass matrix of the right-hnded neutrinos. Thus we would like to see what profile the NO and IO benchmark points for the left-handed neutrinos discussed above infer on the on the right-handed neutrinos.

For the values of the Yukawa couplings associated to the  NO case discussed above, the model predicts, $m_1^R\approx 0$, $m_2^R\approx 19$ eV and $m_3^R\approx 24$ eV, while for the case of  IO, the model predicts  $m_3^R\approx 0$, $m_1^R\approx 1$ eV and $m_2^R\approx 50$ eV. In both cases right-handed neutrinos have masses at eV scale. This is curious because they acquire mass via two-loops process. We can understand this  looking in the mass expression above. In it we do not have the surppresion factors  $V_{CKM}$ and $ \sin(2\theta)$.  Instead we have enhance factors as  $M$ and $m_{d^{\prime}}$.

 Note  that, although  these sterile neutrinos do not mix with the active ones, therefore they do not participate in any   short-baseline, reactor and accelerator neutrinos experiments. However, even in this case, they still  interact with the standard fermions. These interactions are mediated  by the new gauge bosons of the 3-3-1 model and by the leptoquarks. Because they are light neutrinos,  their natural source of constraints are derived from their contributions to cosmological and astrophysical observables as $N_{\mathrm{eff}}$, bounds from BBN and CMB anisotropies, etc. Nevertheless, the Yukawa interactions among leptoquarks and fermions will give contributions to some B-meson decays and also lead to rare Higgs decay. Next we check if the scenario we develop in this section is in agreement with such processes\footnote{The cosmological and astrophysical implications will be investigated elsewhere}.

\section{B-meson physics}
In this section we check if the  parameter space of the Yukawa leptoquarks given in Eqs. (\ref{YNO}) and (\ref{YIO}) together with leptoquarks with mass around 2 TeV are in agreement with the physics of the B-meson. 
\subsection{The contributions to \texorpdfstring{$B \to \mu^+ \mu^-$}{B -> mu+ mu-} decay}

\begin{figure}[t]
\centering
\hspace*{-1cm}\includegraphics[width=0.5\columnwidth]{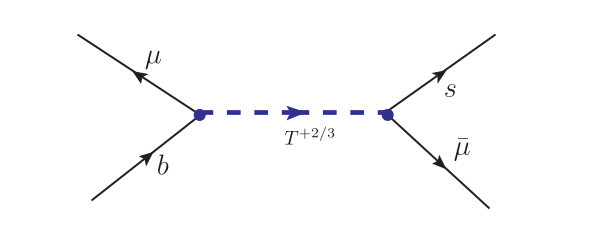}
\vspace*{-0.5cm}
\caption{ Feynman diagram contribution to $B_{s} \to  \mu^+ \mu^-$ decay  for 331LQ model.}
\label{figbs}
\end{figure}

\par The rare decay $B_s \to \mu^+ \mu^-$ is a key process for probing New Physics (NP) at the LHC, as it provides some of the strongest bounds on deviations from the Standard Model (SM) due to its precise theoretical prediction \cite{Bobeth:2013uxa,Altmannshofer:2017wqy}.
 The 331 leptoquarks (331LQ)  contributions to this process can be extract from the Yukawa interaction
\begin{equation}
{\cal L} \supset  Y^{ab}{\bar d}_{aR}  e_{bL} T^{+2/3}  + h.c.
\label{331LQB}
\end{equation}
\noindent and be represented by the diagram showed in the FIG. \ref{figbs}. The scalar leptoquark interactions  via quark-leptoquark penguin diagrams that  potentially contribute to this process in our
model are depicted in  the  FIG.\ref{fighd}.
\par In the SM, the effective Hamiltonian relevant to $b\to s \mu^+ \mu^-$ decay can be written as\cite{Mohanta:2014M}\cite{Sahoo:Sah2015}
 \begin{eqnarray}
 && H_{eff} = - \frac{4G_F}{\sqrt{2}}V_{tb}V^*_{ts}\left[\sum^{{}_{6}}_{{}_{i=1}}C_iO_i +  \frac{\alpha C_7}{4\pi}(\bar{s}\sigma_{\mu\nu}(m_{s}P_L + m_bP_R)b)F^{\mu\nu} + \nonumber\right. \\
 && \left. \hspace*{1cm} +  \frac{\alpha C_9}{4\pi}(\bar{s}\gamma^{\mu}P_L b)(\bar{l}\gamma_\mu l) + \frac{\alpha C_{10}}{4\pi}(\bar{s}\gamma^{\mu}P_L b)(\bar{l}\gamma_\mu \gamma_5 l) \right],  
  \label{HefSM}
\end{eqnarray}
\noindent where in this expression  $V_{tb}V^*_{ts}$ are CKM matrix elements,  $\alpha = \frac{e^2}{4\pi}$ is the fine structure constant,  the chiral projectors are  $P_{L,R} = \frac{1}{2}(1 \pm \gamma_5)$  and  $C_{i}$ to $i=1..10$,  are the Wilson coefficients. Below, we list the operators indicated in Eq.(\ref{HefSM}), as well as the list of other operators susceptible to leptoquarks effects considered in this work
 \begin{eqnarray}
 && O_{7} = \frac{\alpha}{4\pi} (\bar{s}\sigma_{\mu\nu}(m_{s}P_L + m_bP_R)b)F^{\mu\nu} \,\,\,,\,\,\, O_{9} = \frac{ \alpha}{4\pi}(\bar{s}\gamma^{\mu}P_L b)(\bar{l}\gamma_\mu l) \\
 && O_{10} = \frac{\alpha}{4\pi}(\bar{s}\gamma^{\mu}P_L b)(\bar{l}\gamma_\mu \gamma_5 l) \hspace*{1.9cm},\,\,\,O_{S} = \frac{\alpha}{4\pi}(\bar{s}\gamma^{\mu}P_R b)(\bar{l} l) \\
 && O_{P} = \frac{\alpha}{4\pi}(\bar{s}\gamma^{\mu}P_R b)(\bar{l}\gamma_5 l)\hspace*{2.3cm},\,\,\, O_{T} = \frac{\alpha}{4\pi}(\bar{s}\sigma_{\mu\nu} b)(\bar{l}\sigma^{\mu\nu} l) \\
 &&  O_{T5} = \frac{\alpha}{4\pi}(\bar{s}\sigma_{\mu\nu} b)(\bar{l}\gamma_5\sigma^{\mu\nu} l).
  \label{effeO}
\end{eqnarray}

\par The effective Hamiltonian  described by Eq.(\ref{HefSM}) will receive corrections  due to leptoquark interactions  described  in the Eq.(\ref{331LQB}). Considering the Fierz transformation  listed in Appendix \ref{APB}, we can estimate the correction given by the interactions in   Eq. (\ref{331LQB})  to $b\to s \mu^+ \mu^-$ decays.   The effective 331LQ  Hamiltonian that matter is given by 
 \begin{eqnarray}
     && H^{331LQ} =  \frac{Y^{32}Y^{22}}{8M^2_T}(\bar{s}\gamma^{\mu}(1 + \gamma_5) b)(\bar{\mu}\gamma_{\mu}(1 - \gamma_5) \mu) \nonumber \\
     &&  \hspace*{1.35cm} \equiv \frac{Y^{32}Y^{22*}}{4M^2_T}(O_{9'} -O_{10'}),
     \label{H331ef}
\end{eqnarray}
\noindent where we assume for simplicity  $M_T  \approx M_2 $. Note that the set of operators with “primes”,  $O_{9'}$ and $O_{10'}$,   are related to the “unprimed” set by switching the roles of $P_L$ and  $P_R$.
\par  In order to estimate the corrections to the effective operators in the Eq.(\ref{HefSM}), we can write the
 analogous to the Standard Model counterpart in the form 
 \begin{eqnarray}
     && H^{331LQ} =  -\frac{G_F\alpha}{\pi\sqrt{2}}V_{tb}V^*_{ts}(C_{9'}O_{9'} + C_{10'}O_{10'}), 
     \label{H331ef2}
\end{eqnarray}
\noindent  which allows the identification of the new Wilson coefficients $C_{9'}$ and $C_{10'}$ associated with the operators $O_{9'}$ and $O_{10'}$  
\begin{equation}
C_{10'} = -C_{9'} = \frac{Y^{32}Y^{22*}}{2\sqrt{2}G_F\alpha V_{tb}V^*_{ts}}\frac{\pi}{M^2_T}.
     \label{C10}
\end{equation}
\par Assuming the  effective theory given by Eq.(\ref{HefSM}),in the SM  the only operator contributing to   $B_s \to  \mu^+ \mu^-$ amplitude  is the  $O_{10}$\cite{Damir:2012dnfe}, and the expression for the branching fraction of this decay can be write as 
\begin{equation}
\mathrm{Br}^{{}^{\!\!\!\!\!SM}}(B_s \to  \mu^+ \mu^-) =  \frac{N^B}{\pi}m_{B_s}\beta_{\mu}(m^2_{B_s})|C_{10}|^2
     \label{BrSM}
\end{equation}
\noindent where we define 
 \begin{eqnarray}
     && N^B \equiv  \tau_{B_s}(\frac{G_F\alpha|V_{tb}V^*_{ts}|}{4\pi})^2m^2_{\mu}f^2_{B_s}\nonumber \\
     && \beta_{\mu}(m^2_{B_s}) \equiv \sqrt{1 - 4m^2_{\mu}/m^2_{B_s}},
     \label{NB}
\end{eqnarray}
\noindent and in this expression $f_{B_s}$ is the $B_s$ meson decay constant. However, as  discussed in Ref.\cite{Kristof:2012k}, the average time integrated branching ratio $\overline{\mathrm{Br} }(B_s \to  \mu^+ \mu^-)$ depend on the details of $\bar{B_s}-B_s$ mixing,  which in the SM are related to the decay widths $\Gamma(B_s \to  \mu^+ \mu^-)$ by a very simple relation obtained  in these reference,  $\overline{\mathrm{Br}}(B_s \to  \mu^+ \mu^-) = \Gamma(B_s \to  \mu^+ \mu^-)/\Gamma^s_H $, where $\Gamma^s_H$ is the total width of the heaviest mass eigenstate. 
\par The current SM prediction for these branching ratio corresponds to \cite{Czaja:2024M}
\begin{equation}
\overline{\mathrm{Br}}^{{}^{SM}}(B_s \to  \mu^+ \mu^-) = (3.64 \pm 0.12)\times 10^{-9}, 
     \label{BrSM2}
\end{equation}
\noindent and the current world average stands for\cite{PDG:2024PD}
\begin{equation}
\overline{\mathrm{Br}}^{{}^{exp}}(B_s \to  \mu^+ \mu^-) = (3.34 \pm 0.27)\times 10^{-9}, 
     \label{Brexp}
\end{equation}
\par The 331LQ  contributions  to the corresponding branching ratio is given as
\begin{equation}
\mathrm{Br}^{{}^{\!\!\!\!\!331LQ}}(B_s \to  \mu^+ \mu^-) =  \frac{N^B}{\pi}m_{B_s}\beta_{\mu}(m^2_{B_s})|C_{10} - C_{10'}|^2, 
     \label{Br331LQ}
\end{equation}
\noindent taking over corrections discussed in the paragraph above, considering Eqs.(\ref{BrSM}-\ref{Br331LQ})  as well as $C_{10'}$ given be Eq.(\ref{C10}), assuming  the current experimental limit() we obtain the following  constraints on leptoquark couplings as

\begin{equation}
 0 \lesssim  \frac{|Y^{32}Y^{22*}|}{M^2_T}    \lesssim 1.37\times 10^{-9}   \,\mathrm{GeV}^{-2}.
  \label{LQB}
\end{equation}

For leptoquarks at TeV scale ($M_T \approx 1-2$ TeV) we obtain the lower bound $|Y^{32}Y^{22*}| < 1.37 \times 10^{-3}$. Our illustrative case in Eqs. (\ref{YNO}) and (\ref{YIO}) respect such bound. 

\subsection{The contributions to \texorpdfstring{$b_R \to s_R \nu_L \nu_L$}{bR -> sR nuL nuL} decay}

The correlations between $\mathrm{B}(B \to K\nu\nu)$ and $\mathrm{B}(B \to K^*\nu\nu)$ provide important insights into the possible contributions of New Physics (NP) to the effective operators governing these rare decays. Both channels are mediated by the $b \to s\nu\bar{\nu}$ transition, which is highly suppressed in the Standard Model (SM) due to loop-level processes and small effective couplings.

In the SM, the relation between these branching ratios is well-defined, with $\mathrm{B}(B \to K^*\nu\nu)$ expected to be larger due to the additional degrees of freedom from the $K^*$ polarization states. 

The Belle II experiment recently observed the decay $B^+ \to K^+ + \text{inv}$ with a measured branching ratio \cite{BelleII:2021gkm}:
\[
\mathrm{B}(B^+ \to K^+ + \text{inv})_\text{exp} = (2.3 \pm 0.7) \times 10^{-5},
\]
exceeding the SM prediction  \cite{Buras:2014fpa}:
\[
\mathrm{B}(B^+ \to K^+\nu\nu)_\text{SM} = (4.44 \pm 0.30) \times 10^{-6}.
\] This $2.7\sigma$ discrepancy has sparked interest in NP scenarios, including effective field theories and dark sector particles.
From the other side, Belle imposed a superior bound on the decay $B^+ \to K^* + \text{inv}$ at $90\%$ C.L. \cite{Belle:2017oht}. Thus, due to the large correlation between $K$ and $K^\star$, not all effective operators generated by NP can alleviate the tension, and sometimes they could even increase it\cite{Rosauro-Alcaraz:2024mvx}. This is why this section is dedicated to analyzing the correlation between $K$ and $K^\star$ decays into the 331LQ model.

\begin{figure}[h]
\centering
\hspace*{-1cm}\includegraphics[width=0.6\columnwidth]{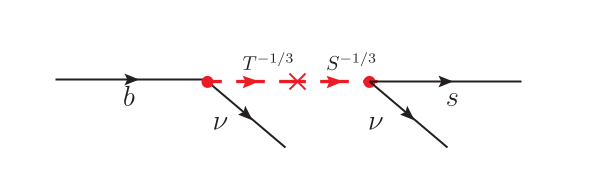}
\vspace*{-0.5cm}
\caption{ Feynman diagram contribution to $b \to  s \nu_L  \bar \nu_L$ decay in 331LQ model.}
\label{figbnu}
\end{figure}

FIG. \ref{figbnu} represents the Feynman diagram of the main contribution  of the 331LQ to the decay of $B$ into $K^{(*)} \nu\nu$.
The interactions in the flavor basis  that lead to this process are
\begin{equation}
Y_{ab}\bar \nu_{b_L} d_{a_R} T^{1/3}+ \tilde Y_{ai} \bar \nu_{a_L} (d_{i_L})^C S^{1/3} +  \frac{M v_\eta}{\sqrt{2}} T^{+1/3} S^{-1/3},
\end{equation}
\noindent where this contribution is represented by the diagram  FIG.\ref{figbnu}. 
In the physical basis we have
\begin{eqnarray}
Y_{ab}\bar{ \hat{\nu}}_{pL}(U_{PMNS})_{pb}  \hat d_{aR}(C_\theta \tilde T^{1/3}-S_\theta \tilde S^{1/3}) + \tilde Y_{ai}\bar{\hat \nu}_{pL}(U_{PMNS})_{ap}  (V^d_L)_{bi}\hat{d_b}_{R}(C_\theta \tilde S^{1/3} + S_\theta \tilde T^{+1/3})
\end{eqnarray}
 From the interaction described above, we get the following effective  operator \begin{eqnarray}
  {\cal M}_{p,p^\prime}^{q^2\to0}=  \sum_{a,a',i=1}^3  \frac{\sin 2{\theta}\, Y_{a3} (V^{d}_L)_{2i} \tilde Y_{a^\prime i}U^{PMNS\,*}_{ p^\prime a^\prime}U^{PMNS}_{a p}(M_2^2-M_1^2)}{8M_1^2M_2^2}(\bar s_R  {\nu_L}_p)(\bar \nu_{Lp^\prime}b_R
  ) ,
    \label{H331net}
\end{eqnarray}
such that the indices $p,p'$ varies between 1 to 3, representing the flavor basis of the neutrinos in the interaction. From these operators, after a Fierz transformation, it is possible to write the relevant effective Hamiltonian of  331LQ contributions to $b \to  s \nu_L  \bar \nu_L$:   \begin{eqnarray}
     && H^{\nu}_{331LQ} =  \frac{G_F\alpha}{2\sqrt{2}\pi}V_{tb}V^*_{ts}\sum_{p,p'}(C_{9^\prime}^{p,p'}O_{9^\prime}^{p,p'} +  C_{10^\prime}^{p,p'}O_{10^\prime}^{p,p'}), 
     \label{H331net1}
\end{eqnarray}
\noindent  with the Wilson coefficients $C_{9^\prime}$ and $C_{10^\prime}$ given by 
\begin{eqnarray}
&& C_{9^\prime}^{p,p'} = \frac{\pi \sin 2{\theta}\, Y_{a3} (V^{d}_L)_{2i} \tilde Y_{a^\prime i}U^{PMNS\,*}_{p^\prime a^\prime}U^{PMNS}_{a p}(M_2^2-M_1^2)}{2\sqrt{2}G_F\alpha V_{tb}V_{ts} M_1^2M_2^2}  = -C_{10^\prime}^{p,p'} \label{neutrino_decay}
\end{eqnarray}
\noindent where  $\theta$ is the angle that rotate  $T^{1/3}$ and $S^{1/3}$  in the physical eigenstates.

\begin{figure}[t]
\centering
\hspace*{-1cm}\includegraphics[width=0.5\columnwidth]{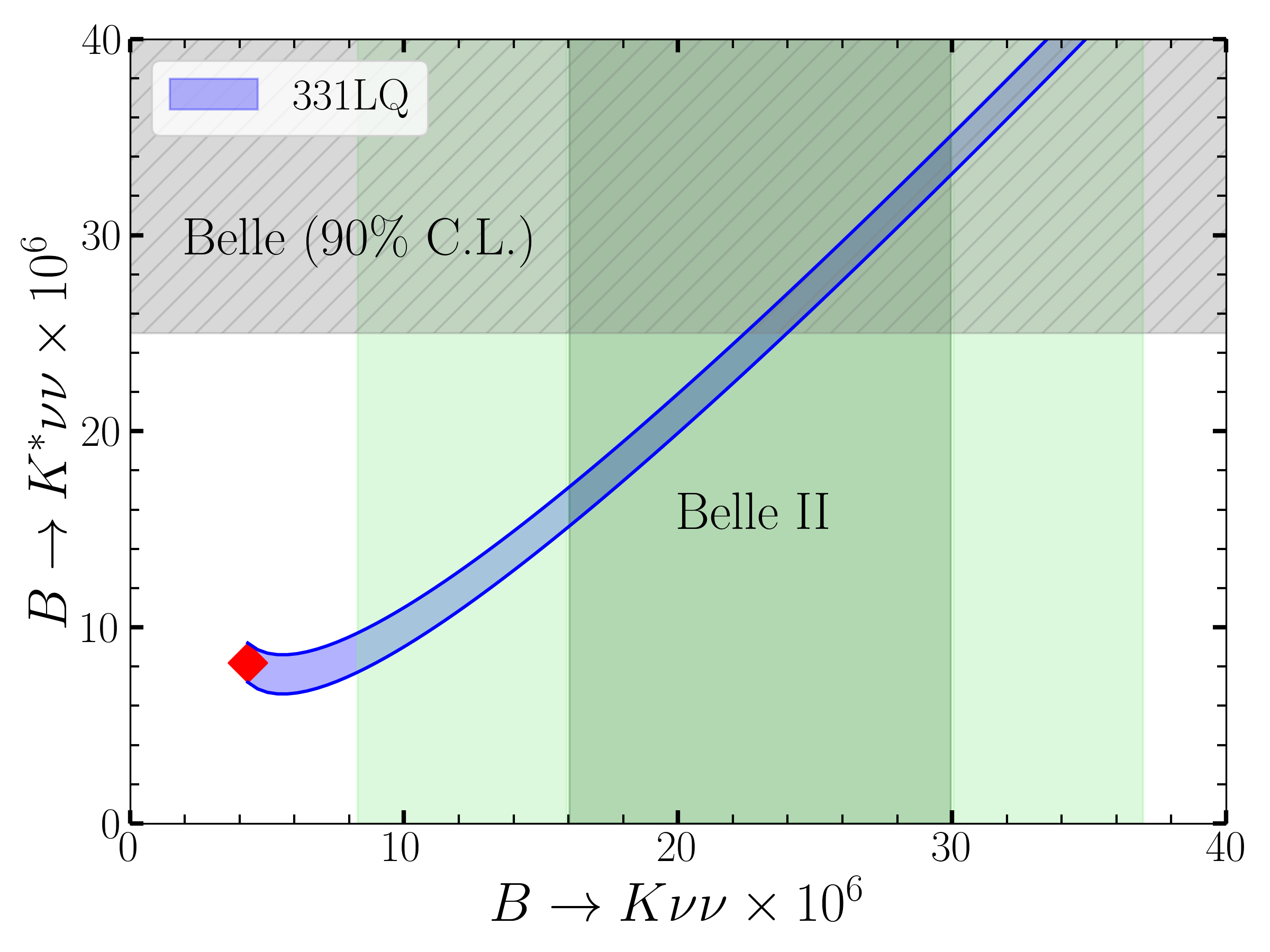}
\vspace*{-0.5cm}
\caption{ Correlation between $\mathrm{B}(B \to K\nu\nu)$ and $\mathrm{B}(B \to K^*\nu\nu)$ when including the effect of the contributions of 331LQ to New Physics (NP) operators using the formula of \cite{Rosauro-Alcaraz:2024mvx}. The red point corresponds to the SM result and the (light) green regions to the Belle II result at 1(2)$\sigma$\cite{BelleII:2021gkm}. The hatched gray areas on the panel correspond to the Belle upper bound, $\mathrm{B}(B \to K^*\nu\nu)< 2.5 \times 10^{-5}$\cite{Belle:2017oht}. } 
\label{fignunu}
\end{figure}

In FIG. \ref{fignunu}, the correlation between $\mathrm{B}(B \to K\nu\nu)$ and $\mathrm{B}(B \to K^*\nu\nu)$ is analyzed, considering the impact of contributions from 331LQ to New Physics (NP) operators using the formula of \cite{Rosauro-Alcaraz:2024mvx}. The red point represents the Standard Model (SM) prediction, while the (light) green regions indicate the Belle II results at $1(2)\sigma$. The hatched gray areas correspond to the Belle upper bound, $\mathrm{B}(B \to K^*\nu\nu) < 2.5 \times 10^{-5}$\cite{Belle:2017oht}. Using the benchmark points for the NO and IO scenarios, along with the previously discussed benchmarks, the contribution to the SM prediction exhibits a negligible correction, on the order of $10^{-5}$. This small correction can be attributed to the fact that $Y_{a3}$ is always of the order of $10^{-3}$, while the dominant term in the product $(V_L^d)_{2i}\tilde{Y}_{a'i}$ is proportional to $10^{-1}$. Consequently, we find that 
\begin{eqnarray}
C_{9'} \approx -10^5 \frac{\sin 2\theta (M_1^2 - M_2^2)}{M_1^2 M_2^2}\to C_{9'} \approx -10^{-2} \sin 2\theta \frac{(1\,\mathrm{TeV})^2}{M_2^2},
\label{c9p}
\end{eqnarray}

while assuming $M_1 = 1.3 M_2$.

For $C_{9'}$ to be relevant, it must satisfy the condition $|C_{9'}| > |C_L^{\mathrm{SM}}| \approx 6$\cite{Rosauro-Alcaraz:2024mvx}, where $C_L^{\mathrm{SM}}$ represents the SM Wilson coefficient for the same process.  According to our finding in Eq. (\ref{c9p}), we have    $C_{9'}< 10^{-2}$, indicating  that  the contribution from leptoquarks alone  does not account for the experimental result observed by   Belle II collaboration. However, the 331RHN includes additional contributions to such processes  mediated by neutral scalars and the  gauge boson $Z^{\prime}$  that, potentially, may accommodate Belle II collaboration result. This will be checked elsewhere.\footnote{Our proposal here is just to check the contributions from   leptoquarks alone in order to check if they enter in conflict with existent experimental results. }

\clearpage

\section{Rare Higgs decay}

\begin{figure}[t]
\centering
\hspace*{-1cm}\includegraphics[width=0.8\columnwidth]{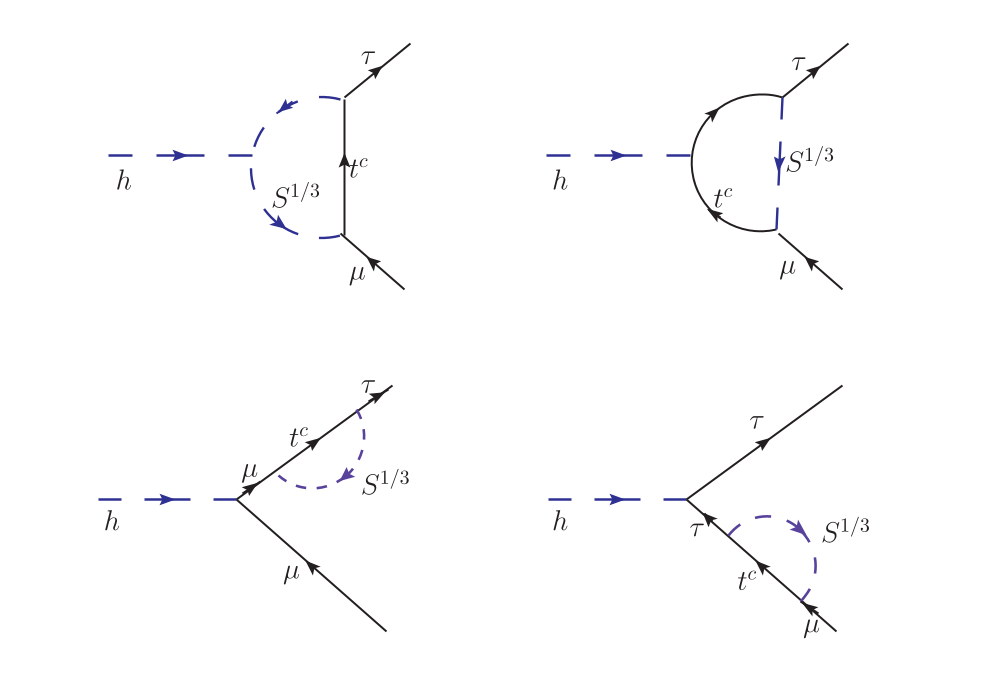}
\vspace*{-0.5cm}
\caption{ Feynman diagrams for 331LQ contributions to $h \to \tau\mu$ decay.}
\label{fighd}
\end{figure}

\par In the Standard Model (SM), flavor-violating (FV) Higgs interactions are highly suppressed \cite{Dorsner:2016wpm,Branco:2011iw}. However, the ATLAS Collaboration has recently reported an excess in the search for the decay $h \to \tau\mu$ \cite{ATLAS:2019old}. In our model, scalar leptoquark interactions can contribute to this process via quark-leptoquark penguin diagrams, as illustrated in FIG.~\ref{fighd}. These diagrams involve a helicity flip on one of the fermion lines, which is essential for generating the amplitude.  
\par The relevant interactions from those presented in Eq.~(\ref{yukawa}) that lead to this process are the Yukawa couplings involving scalar leptoquarks and fermions. These couplings mediate flavor-violating transitions via loop-level diagrams, such as quark-leptoquark penguin diagrams shown in FIG.~\ref{fighd}. In particular, terms where leptoquarks couple to both quarks and leptons play a critical role in inducing the $h \to \tau\mu$ decay. These interactions provide the necessary helicity flip on the fermion line to generate the amplitude for this rare process \cite{Dorsner:2016wpm,Babu:2010vp}

\begin{equation}
  {\cal L} \supset \bar{\hat{u}}^c_{a}(\tilde{Y}_{ia}P_L + y_{ab}P_R)e_b \tilde S^{1/3} +  H.c, 
  \label{RHd1}
\end{equation}
\noindent   After the  explicit computation of the penguin diagrams, that can be found in Refs.\cite{DORSNER:2016RLQ}\cite{DORSNER:2015Ht}, the 331LQ contributions to the effective  $\mu - \tau-h$  coupling take the form
\begin{equation}
 y^{\tau\mu}_{(\mu\tau)} = -\frac{3}{16\pi^2}\frac{m_t}{v}f_{1}(\lambda, M_{2})(\tilde{Y}_{33}y_{32})(\tilde{Y}_{32}y_{33}),
  \label{RHd2}
\end{equation}
\noindent where for reasons of simplification, and without any loss for the model, we consider that charged leptons, neutrinos, right-handed quarks and the new quarks come in a diagonal basis\cite{Doff:2024cap}. 

The Higgs portal coupling, as defined in Ref.\cite{DORSNER:2016RLQ}, is given in our case as,
$$
{\cal L} \supset -(\tilde \lambda_6 + \tilde \lambda_7) v h S^{\dagger}S\equiv-\lambda v h S^{\dagger}S, 
$$
where we have made the usual assumptions: $v_\eta=v_\rho=\frac{v}{\sqrt{2}}$ with $h=\frac{R_\eta}{\sqrt{2}} +\frac{R_\rho}{\sqrt{2}}$ playing the role of the standard Higgs. The function $f_{1}(\lambda, M_{2})$ is the one-loop function given by 
\begin{eqnarray}
 && f_{1}(\lambda, M_{2}) = (M^2_{2} +m^2_t)C_{0}(0,0,m^2_h,m^2_t,M^2_2,m^2_t) + B_{0}(m^2_h,m^2_t,m^2_t) + \nonumber \\
 && \hspace*{1.9cm} - B_{0}(0,m^2_t,M^2_2) + \lambda v^2C_{0}(0,0,m^2_h,M^2_2,m^2_t,M^2_2) ,
  \label{RHd3}
\end{eqnarray}

\noindent such that $(B_{0}, C_{0})$ are the Passarino-Veltman functions, where $B_{0}(0, m_t^2, M_2^2)$ has an analytical expression analogous to Eq.~(\ref{B00PV}). With these components at hand, we now have all the necessary elements to compute the decay width for the $h \to \tau\mu$ process.

 \par The  $h \to \tau\mu$  partial decay width in our model stands for
 \begin{equation}
 \Gamma(h \to \tau\mu) = \frac{9m_hm^2_t}{2^{13}\pi^5v^2}F^2(\tilde{Y})|f_{1}(\lambda, M_{2})|^2,
  \label{RHd4}
\end{equation}
 \noindent such that $F^2(\tilde{Y}) \equiv |\tilde{Y}_{33}y_{32}|^2 + |\tilde{Y}_{32}y_{33}|^2$.

\par In this section, we investigate the contributions from leptoquarks to the process $h \to \tau \mu$. As discussed in Ref.~\cite{PIRES:2024dc}, the introduction of scalar leptoquarks into the 331RHN model provides a solution to the $(g-2)_\mu$ puzzle. By comparing with the experimental results for $(g-2)_\mu$, we derived a bound on the parameter space, specifically $y_{32} \lesssim 0.5$. For the coupling $\tilde{Y}$, we utilize the benchmark points provided in Eqs.~(\ref{YNO}) and (\ref{YIO}) for the normal ordering (NO) and inverted ordering (IO) cases, while the parameter $y_{33}$ remains free, constrained only by the perturbativity limit $(y_{33} \lesssim 4\pi)$. 

\par In addition, we assume $M_{LQ} = 2~\mathrm{TeV}$ and $y_{33} \approx 1$. With these parameters, we predict the branching ratio for the $h \to \tau\mu$ decay as follows:
\begin{equation}
    Br^{\text{331LQ}}(h \to \tau\mu) = 1.56 \times 10^{-4} \, \%.
\end{equation}

\par The current limit established by the ATLAS Collaboration for this branching ratio is \cite{ATLAS:Htmu20,ATLAS:Htmu}:
\begin{equation}
    Br^{\text{ATLAS}}(h \to \tau\mu) = 0.37^{+0.14}_{-0.10} \, \%.
\end{equation}

\par As seen above, the predicted branching ratio $Br^{\text{331LQ}}(h \to \tau\mu)$ is consistent with the current limit established by the ATLAS Collaboration.

\section{Summary and conclusion}
In the original version of the   3331RHN model, neutrinos are massless particles. Numerous attempts have been made to generate small masses for the neutrinos within the 331RHN model. However, the majority of the  attempts are based on seesaw mechanisms. In this paper we  focused on  radiative corrections. A particulary interesting and efficient scenario for  generating radiative neutrino masses is one based on   leptoquarks. We, then, added the most economical set of leptoquarks to the 331RHN that is capable of generating small masses for the standard neutrinos. The leptoquark content encompasses a triplet, $T$, and a singlet, $S$, of scalar leptoquarks. This content attended efficiently our goal and standard neutrinos acquired masses through one-loop corrections and sub-eV neutrino mass was obtained for   leptoquarks at TeV scale.

In the 331RHN model, right-handed neutrinos, along with the standard leptons, form the triplet of leptons. The Yukawa interactions that generate masses for the standard neutrinos also generate masses for the right-handed neutrinos. As interesting results,   right-handed neutrinos acquired masses via two-loops corrections, making them, inexorably, light particles. It is important to emphasize that, even though right-handed neutrinos acquired masses at two-loop level, in our illustrative scenario, considering both NO and IO cases, they still became slightly heavier than the standard neutrinos. This occurred because the masses of these neutrinos are proportional to the heavy particles of the model, which is sufficient to enhance their masses beyond those of the standard neutrinos.

A notable aspect of the 331LQ model concerning neutrino masses is that it does not induce mixing among right-handed and left-handed neutrinos. Consequently right-handed neutrinos are free from  short-baseline neutrinos experimental constraints. Although these right-handed neutrinos are sterile neutrinos, they interact with standard fermions via the new gauge bosons of the 331LQ model rendering them susceptible to cosmological and astrophysical constraints. 

We also examined whether the benchmark points in Eqs. (\ref{YNO}) and (\ref{YIO}) satisfy  $B$-meson observables and rare Higgs decays constrainst. Regarding $B$-meson physics, we verified that the prediction of the model to  $B\to \mu^+ \mu^-$ and the correlations between $\mathrm{B}(B \to K\nu\nu)$ and $\mathrm{B}(B \to K^*\nu\nu)$ decay are consistent with current experimental results. In the Higgs decay inducing LFV,  we verified that the benchmark points in Eqs. (\ref{YNO})  and (\ref{YIO}), in conjuction with ($g-2$) solution, easily  comply the bounds originating from $h \to \tau \mu$ decay imposed by ATLAS collabration.

All our findings  indicate that  the 331LQ model we have presented and developed here  is both  economic and efficient in generating small masses for the neutrinos. Moreover this scenario has as signature leptoquarks with masses at TeV scale that are being probed  currently at the LHC.

\section*{Acknowledgments}
C.A.S.P  was supported by the CNPq research grants No. 311936/2021-0. J.P.P. is supported by grant  PID2022-\allowbreak 126224NB-\allowbreak C21 and  "Unit of Excellence Maria de Maeztu 2020-2023'' award to the ICC-UB CEX2019-000918-M  funded by MCIN/AEI/\allowbreak 10.13039/\allowbreak 501100011033, 
also supported by the European Union's through the
Horizon 2020 research and innovation program (Marie
Sk{\l}odowska-Curie grant agreement 860881-HIDDeN) and the Horizon
Europe research and innovation programme (Marie Sk{\l}odowska-Curie
Staff Exchange grant agreement 101086085-ASYMMETRY). 
It also receives support  from grant 2021-SGR-249 (Generalitat de Catalunya).

\newpage
\appendix
\section{Two-loop parameters}
\label{APA}
\par The function $h$  in Eq.(\ref{MnuR4}) is defined according
\begin{eqnarray}
&& h(M_{d^\prime}, M_{\eta^{\prime}_0},m_{b}, M_2) = 
4\int d^4p\int d^4k\frac{1}{(p^2 - M^2_2)}\frac{1}{(p^2 - m^2_{d^\prime_i})}\nonumber \\
&&\hspace*{4.1cm}\times \frac{1}{(k^2 - M^2_2)}\frac{1}{(k^2 -m^2_{b})}\frac{1}{((p-k)^2 -M^2_{\eta^{\prime}_0})}, 
    \label{int1}
\end{eqnarray}
\noindent after performing a Wick rotation, the integral above can be  written as follows\cite{McDonald:2003MK}

\begin{eqnarray}
&& h(m_{d^\prime_i}, M_{\eta^{\prime}_0},m_{b},M_2) = \frac{8}{(M^2_2 - m^2_{d^\prime})}\frac{1}{(M^2_2 - m^2_{b})}\sum^{{}_{3}}_{{}_{s=1}}f_{s}(m_{d^\prime}, M_{\eta^{\prime}_0}, M_2,m_{b})
    \label{intM}
\end{eqnarray}
\noindent where $m_{d^\prime_i} \equiv m_{d^\prime} $ is the mass of the new  quark d type relevant to the 331RHN mass scale $\mu_{331}$, $m_{b}$ is the bottom quark  mass. 
As we commented all leptoquarks in the FIG.\ref{a331rhn}(b) lie in the TeV scale characterized by
$M_2$. In the expression \ref{intM}  the functions $f_{s}(m_{d^\prime_i}, M_{\eta^{\prime}_0}, M_2,m_{b})$, to $s=1..3$ are 

\begin{eqnarray}
&& f_{1}(m_{d^\prime}, M_{\eta^{\prime}_0}, M_2,m_{b})  = \pi^4 m^2_{d^\prime}\left(F(a_1, b_1) - F(a_2, b_2)\right) + \pi^4 m^2_{b}\left(F(a_3, b_3) - F(a_4, b_4)\right) \nonumber \\
&& f_{2}(m_{d^\prime}, M_{\eta^{\prime}_0}, M_2,m_{b})  = \pi^4 M^2_{\eta^{\prime}_0}\left(F(a_5, b_5) + F(a_6, b_6) - F(a_7, b_7) - F(a_8, b_8) \right)  \nonumber \\
&& f_{3}(m_{d^\prime}, M_{\eta^{\prime}_0}, M_2,m_{b})  = \pi^4 M^2_2\left(F(a_9, b_9) + F(a_{10}, b_{10}) - 2F(a_{11}, b_{11}) \right). 
\label{fint}    
\end{eqnarray}
\noindent and  the  auxiliary functions $F(a_i, b_i)$, with $i=1..11$,   are defined by
\begin{equation}
F(a_i, b_i) = \int^{1}_{0}dx\left(Li_2(1-\mu^2_i) -\frac{\mu^2_i \log(\mu^2_i)}{1-\mu^2_i} \right)   
\label{coef3}
\end{equation} 
\noindent in which
\begin{equation}
\mu_i^2=\frac{a_{i}x + b_{i}(1-x)}{x(1-x)}, 
\label{coef2}
\end{equation}
\noindent and $Li_2(x)$ is the  dilogarithm function  defined as
\begin{equation}
Li_2(x) = -\int^{x}_{0}dy\frac{\log(1-y)}{y}.     
\label{coef4}
\end{equation}
\par The respective $(a_i, b_i)$ coefficients listed in Eq.(\ref{fint}) are defined according to
\begin{eqnarray}
 &&a_1=\left(\frac{\text{$M_2$}}{\text{$m_{d^\prime}$}}\right)^2\,\,,\,\,a_2=\left(\frac{\text{$m_{b}$}}{\text{$m_{d^\prime}$}}\right)^2\,\,,\,\,b_1= b_2=\left(\frac{\text{$M_{\eta^{\prime}_0}$}}{\text{$m_{d^\prime}$}}\right)^2\nonumber\\
&&a_3=\left(\frac{\text{$M_2$}}{\text{$m_{b}$}}\right)^2\,\,,\,\,a_4=\left(\frac{\text{$m_{d^\prime}$}}{\text{$m_{b}$}}\right)^2\,\,,\,\,b_3=b_4=\left(\frac{\text{$M_{\eta^{\prime}_0}$}}{\text{$m_{b}$}}\right)^2\nonumber\\
&&a_5=\left(\frac{\text{$M_2$}}{\text{$M_{\eta^{\prime}_0}$}}\right)^2\,\,,\,\,b_5=\left(\frac{\text{$m_{b}$}}{\text{$M_{\eta^{\prime}_0}$}}\right)^2\!\!\,\,,\,\,b_6=a_8=b_8=\left(\frac{\text{$M_2$}}{\text{$M_{\eta^{\prime}_0}$}}\right)^2\nonumber\\
&&a_9=\left(\frac{\text{$m_{b}$}}{\text{$M_2$}}\right)^2\,\,,\,\,b_7=\left(\frac{\text{$m_{b}$}}{\text{$M_{\eta^{\prime}_0}$}}\right)^2\,\,,\,\,b_9 = b_{10} = b_{11}=\left(\frac{\text{$M_{\eta^{\prime}_0}$}}{\text{$M_2$}}\right)^2\nonumber\\
&&a_{10} = \left(\frac{\text{$m_{d^\prime}$}}{\text{$M_2$}}\right)^2\,\,,\,\, a_6=a_7=\left(\frac{\text{$m_{d^\prime}$}}{\text{$M_{\eta^{\prime}_0}$}}\right)^2\,\,,\,\,a_{11}=1.  
\label{coef}
\end{eqnarray}

 \section{Fierz transformation}
 \label{APB}
\par The effective operator $(\bar{\chi}_R\Psi_L)(\bar{\Psi}_L\chi_R)$ can be written in the form 
\begin{equation}
(\bar{\chi}_R\Psi_L)(\bar{\Psi}_L\chi_R) =  \frac{1}{4}((\bar{\chi}\Psi)(\bar{\Psi}\chi) - (\bar{\chi}\gamma_5\Psi)(\bar{\Psi}\gamma_5\chi))   
\label{A21}
\end{equation}
\par Assuming  the  product of  bilinear spinors decomposed in the Fierz base 
according to
\begin{eqnarray}
&& (\bar{\chi}\Psi)(\bar{\Psi}\chi) = \frac{1}{4}(\bar{\chi}\chi)(\bar{\Psi}\Psi) + \frac{1}{4}(\bar{\chi}\gamma^{\mu}\chi)(\bar{\Psi}\gamma_{\mu} \Psi) + \frac{1}{8}(\bar{\chi}\eta^{\mu\nu}\chi)(\bar{\Psi}\eta_{\mu\nu} \Psi) + \nonumber \\
&& \hspace*{2.1cm} - \frac{1}{4}(\bar{\chi}\gamma^{\mu}\gamma_5\chi)(\bar{\Psi}\gamma_{\mu}\gamma_ \Psi) + \frac{1}{4}(\bar{\chi}\gamma_5\chi)(\bar{\Psi}\gamma_5 \Psi)
    \label{A22}
\end{eqnarray}
\begin{eqnarray}
&& (\bar{\chi}\gamma_5\Psi)(\bar{\Psi}\gamma_5\chi) = \frac{1}{4}(\bar{\chi}\chi)(\bar{\Psi}\Psi) - \frac{1}{4}(\bar{\chi}\gamma^{\mu}\chi)(\bar{\Psi}\gamma_{\mu} \Psi) + \frac{1}{8}(\bar{\chi}\eta^{\mu\nu}\chi)(\bar{\Psi}\eta_{\mu\nu} \Psi) + \nonumber \\
&& \hspace*{2.8cm} + \frac{1}{4}(\bar{\chi}\gamma^{\mu}\gamma_5\chi)(\bar{\Psi}\gamma_{\mu}\gamma_5 \Psi) + \frac{1}{4}(\bar{\chi}\gamma_5\chi)(\bar{\Psi}\gamma_5 \Psi)
    \label{A23}
\end{eqnarray}
\par Considering Eqs.(\ref{A22}) and (\ref{A23}) we can write Eq.(\ref{A21}) as
\begin{eqnarray}
&& (\bar{\chi}_R\Psi_L)(\bar{\Psi}_L\chi_R) =  \frac{1}{8}((\bar{\chi}\gamma^{\mu}\chi)(\bar{\Psi}\gamma_\mu\Psi) - (\bar{\chi}\gamma^{\mu}\gamma_5\chi)(\bar{\Psi}\gamma_{\mu}\gamma_5\Psi))   \nonumber\\
&&\hspace*{2.9cm} = \frac{1}{8}((\bar{\chi}\gamma^{\mu}(1 + \gamma_5)\chi)(\bar{\Psi}\gamma_\mu(1 - \gamma_5)\Psi).
\label{A24}
\end{eqnarray}
\par  The Fierz transformations are typically straightforward for orthogonal operators, like scalar-scalar and vector-vector operators. However, for non-orthogonal operators, such as scalar-pseudo scalar  operators, the transformation is more complex due to the different nature of these operators. In these case we have 
 \begin{eqnarray}
&& (\bar{\chi}\Psi)(\bar{\Psi}\gamma_5 \chi) =  \frac{1}{4}(\bar{\chi}\chi)(\bar{\Psi}\gamma_5 \Psi) + \frac{1}{4}(\bar{\Psi}\Psi)(\bar{\chi}\gamma_5 \chi)
+ \frac{1}{4}(\bar{\chi}\gamma^{\mu}\chi)(\bar{\Psi}\gamma_{\mu}\gamma_5 \Psi) - 
 \frac{1}{4}(\bar{\chi}\gamma^{\mu}\gamma_5\chi)(\bar{\Psi}\gamma_{\mu}\Psi) \nonumber \\
&& \hspace*{2.4cm} + \frac{1}{8}(\bar{\chi}\eta^{\mu\nu}\chi)(\bar{\Psi}\eta_{\mu\nu}\gamma_5 \Psi). 
\label{A25}
\end{eqnarray}


\bibliography{references}

\end{document}